\documentclass[aps,preprint,nofootinbib,superscriptaddress]{revtex4}
\usepackage[dvips]{graphicx}
\usepackage{amsmath}
\usepackage{amssymb}
\usepackage{graphicx}
\usepackage{mathrsfs}
\usepackage{ulem}
\usepackage{times}
\usepackage{dashrule}
\usepackage{multirow}
\makeatletter
\def\bib@device#1#2{}
    
    \newcommand{\Rmnum}[1]{\expandafter\@slowromancap\romannumeral #1@}
  \makeatother

\graphicspath{%
    {img/}
    {converted_graphics/}
}
\begin{document}
\title{Prediction of semi-metallic tetragonal $\mathrm{Hf_2O_3}$ and $\mathrm{Zr_2O_3}$ from first-principles}
\author{Kan-Hao Xue}
\affiliation{IMEP-LAHC, MINATEC-INPG, 3 rue Parvis Louis N\'eel, BP 257, 38016 Grenoble Cedex 1, France}
\author{Philippe Blaise}
\affiliation{CEA-LETI, MINATEC, 17 rue des Martyrs, 38054 Grenoble Cedex 9, France}
\author{Leonardo R. C. Fonseca}
\affiliation{Center for Semiconductor Devices, State University of Campinas, Campinas, SP, Brazil}
\author{Yoshio Nishi}
\affiliation{Department of Electrical Engineering, Stanford University, Stanford, CA 94305, USA}
\begin{abstract}
A tetragonal phase is predicted for $\mathrm{Hf_2O_3}$ and $\mathrm{Zr_2O_3}$ using density functional theory. Starting from atomic and unit cell relaxations of substoichiometric monoclinic $\mathrm{HfO_2}$ and $\mathrm{ZrO_2}$, such tetragonal structures are only reached at zero temperature by introducing the oxygen vacancy pair with the lowest formation energy. The tetragonal $\mathrm{Hf_2O_3}$ and $\mathrm{Zr_2O_3}$ structures belong to space group $P\mathrm{\bar{4}}m\mathrm{2}$ and are more stable than their corundum structure counterparts. These phases are semi-metallic, as confirmed through further $\mathrm{G_0W_0}$ calculations. The carrier concentrations are estimated to be $1.77\times{10^{21}}$\,$\mathrm{cm^{-3}}$ for both electrons and holes in tetragonal $\mathrm{Hf_2O_3}$, and $1.75\times{10^{21}}$\,$\mathrm{cm^{-3}}$ for both electrons and holes in tetragonal $\mathrm{Zr_2O_3}$. The tetragonal $\mathrm{Hf_2O_3}$ phase is probably related to the low resistivity state of hafnia-based resistive random access memory (RRAM).
\end{abstract}
\maketitle
Hafnia ($\mathrm{HfO_2}$) and zirconia ($\mathrm{ZrO_2}$) are found in a number of important technological applications.\textsuperscript{\onlinecite{Lange,Kilner}} In particular, hafnia has become a key component in sub-micrometer silicon MOS technology as the current choice of high permittivity dielectric layer.\textsuperscript{\onlinecite{Robertson2006}} In addition, it is also a promising candidate material for resistive random access memory (RRAM), which is one of the leading technologies for the next-generation non-volatile memory.\textsuperscript{\onlinecite{Asamitsu,Ignatiev,Waser2007,Kope}} The core element of RRAM is a metal/insulator/metal capacitor which is subject to an electroforming process, where a high electric field (some MV/cm) is applied across the capacitor to create conduction paths in the insulating thin film, here named filaments. These filaments are of unknown composition or shape, and can be easily disturbed under electrical stress, leading to a memory effect. Knowing the composition of the filaments is crucial to the understanding of RRAM's physics. Previous work on $\mathrm{TiO_2}$ RRAM reveals that the conductive filament is possibly due to $\mathrm{Ti_nO_{2n-1}}$ Magn\'eli phases, where the value of $n$ is mostly 4 or 5.\textsuperscript{\onlinecite{Ti4O7}} For hafnia RRAM, the structure of the conductive filaments has not been reported, though it is widely accepted that the filaments are associated to an oxygen-deficient phase.\textsuperscript{\onlinecite{Lin2011,Kamiya}} Since the impact of electroforming is expected to occur in small and random patches of the capacitor, experimental investigation of the filaments suffers from great difficulty.

In the present paper we employ first-principles density functional theory\textsuperscript{\onlinecite{DFT}} (DFT) calculations to identify O-poor stable compositions of hafnium and zirconium oxides which may be reachable from the room temperature normal pressure monoclinic $\mathrm{Hf(Zr)O_2}$  [$m$-$\mathrm{Hf(Zr)O_2}$] with the assistance of an external source of energy, possibly an applied electric field. The processing and operation of oxide-based RRAM (OXRRAM) stimulated our search of a conductive phase in these materials. However, our predictions are quite general and may have broader implications.

For DFT calculations, plane-wave based Vienna Ab initio Simulation Package (VASP) program\textsuperscript{\onlinecite{VASP}} was implemented, using projector-augmented-wave (PAW) pseudopotentials\textsuperscript{\onlinecite{PAW}} with Hf $5p$, $5d$ and $6s$ (Zr $4s$, $4p$, $4d$ and $5s$) electrons and O $2s$ and $2p$ electrons in the valence. Generalized gradient approximation (GGA) was used for the exchange-correlation energy, within the Perdew-Burke-Ernzerhof (PBE) functional.\textsuperscript{\onlinecite{PBE}} The plane wave energy cutoff is chosen as 500\,eV which converges for all the involved compounds, and sufficiently dense Monkhorst-Pack\textsuperscript{\onlinecite{Monkhorst}} or $\Gamma$-centered $k$-mesh was utilized for sampling the Brillouin zone.

Because DFT/GGA usually underestimates band gaps, and may even deem a material metallic rather than semiconductor as the case of bulk germanium,\textsuperscript{\onlinecite{Germanium}} for the metallic or semi-metallic candidates we calculated the first order energy shifts with $\mathrm{GW}$ approximation\textsuperscript{\onlinecite{GW}} ($\mathrm{G_0W_0}$) using the ABINIT\textsuperscript{\onlinecite{Abinit,VASP-Abinit}} program. Hf~(Zr) semi-core electrons were explicitly included through the $5s^25p^65d^26s^2$ ($4s^24p^64d^25s^2$) configuration while core electrons were replaced by Troullier-Martins pseudopotentials.\textsuperscript{\onlinecite{TMPP}} Convergence was achieved with 360 bands and a 15 Ha cutoff for the wave functions employed in the evaluation of the dielectric function and the $\Sigma$ function. A plasmon-pole approximation\textsuperscript{\onlinecite{Plasmon}} was used.

The $m$-$\mathrm{Hf(Zr)O_2}$ unit cells were fully relaxed until all Hellmann-Feynman forces were less than 0.01\,eV/\AA\ and all stresses were less than 400\,MPa. The relaxed $m$-$\mathrm{Hf(Zr)O_2}$ unit cell parameters are a=5.146\,(5.219)\,\AA, b/a=1.010\,(1.012), c/a=1.036\,(1.036) and $\beta=99.68^{\mathrm{o}}$\,($99.68^{\mathrm{o}}$), close to experimental values:\textsuperscript{\onlinecite{Exp-Lattice}} a=5.117\,(5.151)\,\AA, b/a=1.011\,(1.010), c/a=1.034\,(1.032) and $\beta=99.22^{\mathrm{o}}$\,($99.20^{\mathrm{o}}$). The formation enthalpy of $m$-$\mathrm{Hf(Zr)O_2}$ calculated with respect to $h.c.p.$ Hf (Zr) and an isolated $\mathrm{O_2}$ molecule is -1166\,(-1106)\,kJ/mol, after adopting the 1.36\,eV energy correction for the $\mathrm{O_2}$ molecule provided by Wang \textit{et al}.\textsuperscript{\onlinecite{Ceder2006}} These results are in accord with the experimental values, which are -1145\,kJ/mol for hafnia and -1101\,kJ/mol for zirconia.\textsuperscript{\onlinecite{Speight}}
\begin{figure}[!b]
  \centering
  \includegraphics[bb=0 0 586 332,width=3.5in,height=1.98in,keepaspectratio]{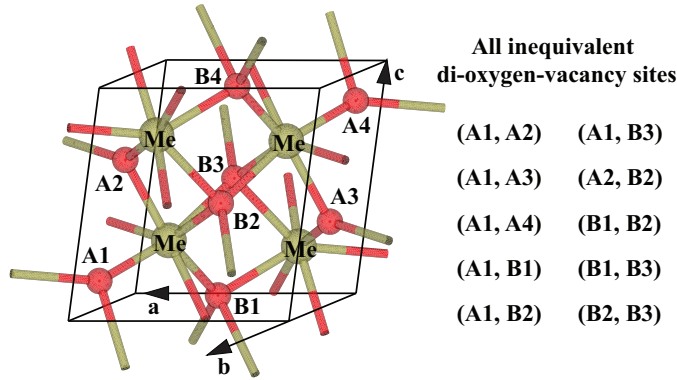}
  \caption{Unit cell used for calculations of the energy of formation of oxygen vacancies in $\mathrm{MeO_2}$ (Me\,=\,Hf or Zr) with all metal and oxygen sites identified. The ten inequivalent pairs of oxygen sites are also listed.}
\end{figure}

For reference we first calculated the formation energies of single and double oxygen vacancies in a 96-atom $m$-$\mathrm{Hf(Zr)O_2}$ supercell ($2\times2\times2$). Aiming at energetically favorable metallic phases, only the neutral oxygen vacancy was considered. There are two inequivalent O sites regarding O coordination, namely the 3- and 4-coordinated O(A) and O(B) sites, respectively. The formation energy of a neutral oxygen vacancy is defined as 
\begin{equation}
E_{\mathrm{for}} = E_\mathrm{D} - E_\mathrm{0} + \mu_\mathrm{O}
\end{equation}
where $E_\mathrm{D}$ is the energy of the defective supercell, $E_\mathrm{0}$ is the energy of the defect-free supercell and $\mu_\mathrm{O}$ is the chemical potential of oxygen. Under oxygen-rich condition $\mu_\mathrm{O}$ is usually set as one half of that of an $\mathrm{O_2}$ molecule. In this case the calculated $E_{\mathrm{for}}$ for a neutral O(A) vacancy in $m$-$\mathrm{Hf(Zr)O_2}$ is 7.13\,(6.62)\,eV, while for a neutral O(B) vacancy it is 7.00\,(6.53)\,eV. The difference between the two formation energies, 0.13\,(0.09)\,eV, is similar to Zheng \textit{et al.},\textsuperscript{\onlinecite{Ceder2007}} but larger than Foster \textit{et al.} who reported a 0.02\,eV difference in both cases.\textsuperscript{\onlinecite{Foster2001,Foster2002}} Next we calculated the formation energies of di-oxygen-vacancy pairs in $m$-$\mathrm{Hf(Zr)O_2}$. To this end, the 8 oxygen sites in a unit cell were named A1--A4 and B1--B4, as shown in Fig.~1. The distance between two di-oxygen-vacancy pairs is around 10\,\AA, casting them as isolated pairs. Since $m$-$\mathrm{Hf(Zr)O_2}$ possesses the baddeleyite structure with space group $P\mathrm{2}_\mathrm{1}/c$, there are 10 inequivalent di-oxygen-vacancy pairs. The most energetically favorable pair is the (B1,\,B2) pair, whose formation energy per vacancy is the same as of a single O(B) vacancy.

Under usual experimental conditions the dielectric is placed between two metal electrodes where oxygen can migrate as an interstitial. We thus calculated the incorporation energies of oxygen, starting from its molecular form, into bulk $h.c.p.$ Ti and $f.c.c.$ Pt, two commonly used electrodes. The results are -6.24 eV and 0.91 eV, respectively. The formation energies of single and double oxygen vacancy in all three environments are thus compared in Tab.~\Rmnum{1}. Notice that the formation energy of a neutral O(B) vacancy in $m$-$\mathrm{Hf(Zr)O_2}$ plus an oxygen interstitial in $h.c.p.$ Ti is merely 0.76\,(0.29)\,eV, which is attributed to the strong Ti--O bonding.
\begin{table}[!b]
\centering
\begin{tabular}{c c c c c c}
\hline
Chemical & Vacancy & Supercell & \multicolumn{3}{c}{Formation energy (eV/vacancy)} \\
formula & site(s) & units & \multicolumn{3}{c}{with oxygen going to} \\
& & & $\mathrm{O_2}$ & Ti & Pt \\
\hline
& & \multicolumn{2}{c}{Single oxygen vacancy} & & \\
\cline{3-4}
$\mathrm{Me_{32}O_{63}}$ & O(A) & 2$\times$2$\times$2 & 7.13(6.62) & 0.89(0.38) & 8.04(7.53)  \\
$\mathrm{Me_{32}O_{63}}$ & O(B) & 2$\times$2$\times$2 & 7.00(6.53) & 0.76(0.29) & 7.91(7.44) \\
& & \multicolumn{2}{c}{Closest di-oxygen-vacancy} & \\
\cline{3-4}
$\mathrm{Me_{32}O_{62}}$ & (B1, B2) & 2$\times$2$\times$2 & 7.01(6.52) & 0.77(0.28) & 7.92(7.43) \\
$\mathrm{Me_{32}O_{62}}$ & (B2, B3) & 2$\times$2$\times$2 & 7.02(6.60) & 0.78(0.36) & 7.93(7.51) \\
$\mathrm{Me_{32}O_{62}}$ & (A1, B1) & 2$\times$2$\times$2 & 7.03(6.53) & 0.79(0.29) & 7.94(7.44) \\
& & \multicolumn{2}{c}{$\mathrm{Me_{4}O_{7}}$} & \\
\cline{3-4}
$\mathrm{Me_{4}O_{7}}$ & O(A) & 1$\times$1$\times$1 & 7.16(6.66) & 0.92(0.42) & 8.07(7.57) \\
$\mathrm{Me_{4}O_{7}}$ & O(B) & 1$\times$1$\times$1 & 7.03(6.57) & 0.79(0.33) & 7.94(7.48) \\
& & \multicolumn{2}{c}{$\mathrm{Me_{2}O_{3}}$} & \\
\cline{3-4}
$\mathrm{Me_{4}O_{6}}$ & (B1, B2) & 1$\times$1$\times$1 & 6.55(5.66) & 0.31(-0.58) & 7.46(6.57) \\
$\mathrm{Me_{4}O_{6}}$ & (A1, B1) & 1$\times$1$\times$1 & 6.92(6.40) & 0.68(0.16) & 7.83(7.31) \\
$\mathrm{Me_{4}O_{6}}$ & (B2, B3) & 1$\times$1$\times$1 & 6.95(6.33) & 0.71(0.09) & 7.86(7.24) \\
\hline
\end{tabular}
\caption{Formation energies of isolated single oxygen vacancies and di-oxygen-vacancy pairs, of isolated oxygen vacancy chains, and of substoichiometric $\mathrm{Hf(Zr)_{4}O_{7}}$ and $\mathrm{Hf(Zr)_{2}O_{3}}$ in different structures derived from monoclinic $\mathrm{HfO_2}$ and $\mathrm{ZrO_2}$. In the table, Me stands for Hf or Zr; the parenthesed formation energy values are Zr-based while others are Hf-based.}
\end{table}
\begin{figure}[!t]
  \centering
  \includegraphics[bb=0 0 505 409,width=3.5in,height=2.84in,keepaspectratio]{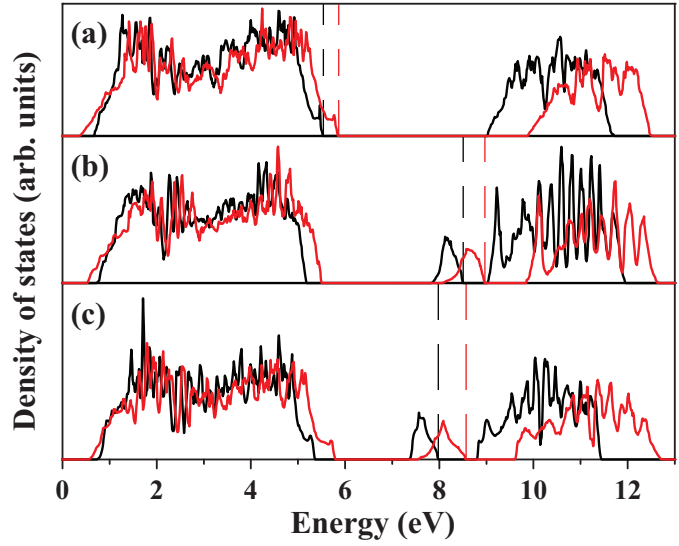}
  \caption{Density of states for (a) $m$-$\mathrm{Hf(Zr)O_2}$; (b) $\mathrm{Hf(Zr)_4O_7}$ with one O(A) vacancy; (d) $\mathrm{Hf(Zr)_4O_7}$ with one O(B) vacancy. Black curves are for Zr-based compounds while red curves are for Hf-based compounds. All figures are aligned horizontally with respect to the O $2s$ band (not shown). The highest occupied molecular orbital levels are indicated by vertical dashed lines.}
\end{figure}

The search for metallic states began with the $\mathrm{Hf(Zr)_4O_7}$ models generated by introducing one oxygen vacancy per 12-atom $m$-$\mathrm{Hf(Zr)O_2}$ for two inequivalent cases, \textit{i.e.}, O(A) and O(B). The defective unit cells were fully relaxed until all Hellmann-Feynman forces were less than 0.01 eV/\AA. Atomic coordinates, cell dimensions and shape were subject to relaxation. The formation energies per vacancy of the two $\mathrm{Hf(Zr)_4O_7}$ phases are almost the same as in the single oxygen vacancy cases. The resulting DOS are shown in Figs.~2(b) and 2(c). In either case, a fully occupied defect-induced band emerges in the band gap, indicating a semiconductor. While the DOS of this band is broad and high, it cannot account for the measured metallic state in hafnia-based RRAM, because in that state the resistance is on the order of hundreds of Ohms for 10\,nm thin films.\textsuperscript{\onlinecite{Leti2011}} However, the trend does hint that stronger off-stoichiometric hafnia or zirconia might undergo a phase transition from dielectric to metal.
\begin{table}[!b]
\centering
\begin{tabular}{c c c c c c}
\hline
 & $a$ & $c$ & $\mathrm{Hf_z}$ & $\mathrm{O(A)_z}$ & Bulk modulus\\
\hline
$\mathrm{Hf_2O_3}$ & 3.135\,\AA & 5.646\,\AA & 0.2553 & 0.1351 & 246\,GPa \\
$\mathrm{Zr_2O_3}$ & 3.174\,\AA & 5.763\,\AA & 0.2525 & 0.1367 & 228\,GPa \\
\hline
\end{tabular}
\caption{Calculated structural parameters of tetragonal $\mathrm{Hf_2O_3}$ and $\mathrm{Zr_2O_3}$.}
\end{table}

Hence, several $\mathrm{Hf(Zr)_2O_3}$ models were set up with two oxygen vacancies per 12-atom $m$-$\mathrm{Hf(Zr)O_2}$ unit cell. For all inequivalent cases (Fig.~1) the cells remain monoclinic during relaxation of the unit cell vectors, except for the (B1,\,B2) case which suffers from a monoclinic-to-tetragonal transition. The tetragonal $\mathrm{Hf(Zr)_2O_3}$ [$t$-$\mathrm{Hf(Zr)_2O_3}$] phase (Fig.~3; structural parameters in Tab.~\Rmnum{2}) is the ground state of all ten $\mathrm{Hf(Zr)_2O_3}$ candidates. It belongs to the $D_{\mathrm{2}d}$ point group and $P\mathrm{\bar{4}}m\mathrm{2}$ (No.~115) space group\textsuperscript{\onlinecite{Aroyo}}. The Hf(Zr) coordination number is 7 as in $m$-$\mathrm{Hf(Zr)O_2}$, while 2/3 of the oxygen sites have coordination number 5 [named O(A)] and 1/3 have coordination number 4 [named O(B)]. Symmetry analysis indicates that Hf(Zr) and O(A) are at the $2g$ position while O(B) is at the $1c$ position\textsuperscript{\onlinecite{Bilbao}}. The average Hf(Zr)--O(A) and Hf(Zr)--O(B) bond lengths are 2.295\,(2.322)\,\AA\ and 2.089\,(2.134)\,\AA, respectively, compared with 2.084\,(2.117)\,\AA\ and 2.209\,(2.240)\,\AA\ in $m$-$\mathrm{Hf(Zr)O_2}$. Bader analysis reveals that the Hf (Zr) charge changes from 2.73e\,(2.57e) in $m$-$\mathrm{Hf(Zr)O_2}$ to 2.10e\,(2.02e) in $t$-$\mathrm{Hf(Zr)_2O_3}$; the O(A) charge changes from -1.34e\,(-1.25e) to -1.39e\,(-1.34e); and the O(B) charge changes from -1.39e\,(-1.31e) to -1.41e\,(-1.37e).
\begin{figure}[!t]
  \centering
  \includegraphics[bb=0 0 483 441,width=4in,height=3.65in,keepaspectratio]{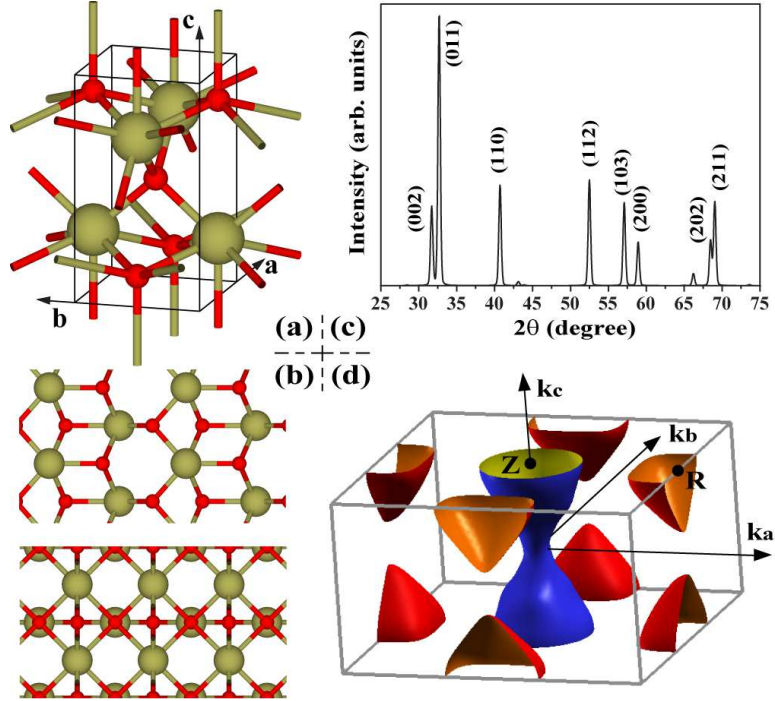}
  \caption{Tetragonal $\mathrm{Hf_2O_3}$ and $\mathrm{Zr_2O_3}$.\textsuperscript{\onlinecite{VESTA}} (a) Primitive cell with 5 atoms; (b) view of the structure along $a$-axis (upper) and $c$-axis (lower); (c) simulated powder X-ray diffraction patterns of $\mathrm{Hf_2O_3}$; (d) Fermi surface of $\mathrm{Hf_2O_3}$ at T=0\,K. Big green balls: Hf or Zr; small red balls: O.}
\end{figure}
\begin{figure}[!b]  \centering
  \includegraphics[bb=0 0 563 387,width=4in,keepaspectratio]{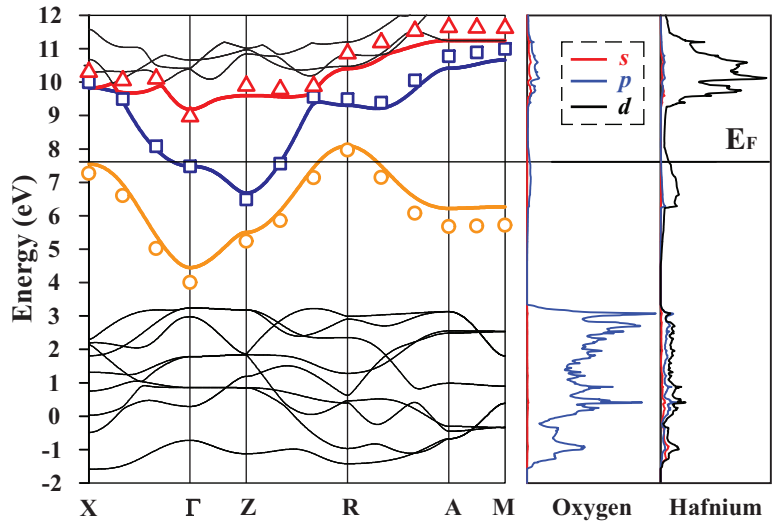}
  \caption{Tetragonal $\mathrm{Hf_2O_3}$ electronic band structure (left) and orbital projected DOS (right). DFT/GGA (solid lines) and $\mathrm{G_0W_0}$ (discrete marks) band structures are superimposed and aligned by their Fermi levels. Results for $\mathrm{Zr_2O_3}$ (not shown) are similar.}
\end{figure}

Figure 4 shows the band diagram and orbital-projected DOS of $t$-$\mathrm{Hf_2O_3}$. Results for $t$-$\mathrm{Zr_2O_3}$ (not shown) are similar. The high symmetry points in the Brillouin zone are named according to the Bilbao crystallographic server.\textsuperscript{\onlinecite{Aroyo}} A semi-metallic behavior is revealed in the overlap of the partially occupied valence band top and conduction band bottom located at different high symmetry points, R and Z. Both band edges are mostly derived from Hf $5d$ states. The semi-metallic character of the two compounds was confirmed by $\mathrm{G_0W_0}$ calculations of the many-body correction to the DFT/GGA energy levels. Figure 4 implies that the energy shifts are of the order of a few tenths of eV, retaining the semi-metallic nature of $t$-$\mathrm{Hf_2O_3}$. The energy shifts obtained for $\mathrm{Zr_2O_3}$ (not shown) are similar. From the calculated band structures the densities of conduction electrons and holes can be obtained by integrating the electron occupation of the blue and orange bands in Fig.~4, respectively. The electron and hole densities are both $1.77\times{10^{21}}$\,$\mathrm{cm^{-3}}$ for $t$-$\mathrm{Hf_2O_3}$, and both $1.75\times{10^{21}}$\,$\mathrm{cm^{-3}}$ for $t$-$\mathrm{Zr_2O_3}$, typical of semi-metals. 

To evaluate the relative stability of this structure, we calculated the molar formation enthalpy of various $\mathrm{Hf_2O_3}$, $\mathrm{Zr_2O_3}$ and $\mathrm{Ti_2O_3}$ models with respect to their corresponding metals and $\mathrm{O_2}$. Still, the 1.36\,eV energy correction to $\mathrm{O_2}$ was applied to all cases. The formation enthalpy of $t$-$\mathrm{Hf(Zr)_2O_3}$ is -1700\,(-1666)\,kJ/mol, more favorable than fully relaxed corundum $\mathrm{Hf(Zr)_2O_3}$, -1586\,(-1580)\,kJ/mol. Nevertheless, a fully relaxed $\mathrm{Ti_2O_3}$ arranged in the $P\mathrm{\bar{4}}m\mathrm{2}$ tetragonal structure, possesses a formation enthalpy of -1576\,kJ/mol, less favorable than corundum $\mathrm{Ti_2O_3}$ whose formation enthalpy is -1598\,kJ/mol. These data confirm that for $\mathrm{Hf(Zr)_2O_3}$ the tetragonal $P\mathrm{\bar{4}}m\mathrm{2}$ structure is preferred, while for $\mathrm{Ti_2O_3}$ the corundum structure is preferred.        
  
To our best knowledge, the proposed $t$-$\mathrm{Hf_2O_3}$ and $t$-$\mathrm{Zr_2O_3}$ structures have not been reported before, though some published data may suggest their existance. Hildebrandt \textit{et al.}\textsuperscript{\onlinecite{Frankfurt}} performed high-resolution transmission electron microscopy of a conducting $\mathrm{HfO_{2-x}}$ thin film where the enlarged inverse Fourier-transformed images show a similar structure as in Fig.~3(b). Manory \textit{et al.}\textsuperscript{\onlinecite{Osaka}} discovered two unidentified X-ray diffraction (XRD) peaks at $2\theta=40^{\mathrm{o}}$ and $2\theta=52^{\mathrm{o}}$ in hafnia films grown by ion beam assisted deposition at a transport ratio of 5 and an ion energy of 20\,keV. They attributed these peaks to a new tetragonal structure and suggested the $\mathrm{Hf_2O_3}$ stoichiometry. However, they simulated their data with a $P\mathrm{4}/mmm$ phase with lattice parameters a=5.055\,\AA\ and c=5.111\,\AA, resulting in two small peaks around $40^{\mathrm{o}}$. We calculated powder XRD patterns\textsuperscript{\onlinecite{XRD}} for $t$-$\mathrm{Hf_2O_3}$ [Fig.~ 3(d)] using Cu $K\alpha$ radiation ($\lambda=1.5418$\,\AA), and found a (110) peak at $40.7^{\mathrm{o}}$ and a (112) peak at $52.5^{\mathrm{o}}$, similar to data. Similar calculation for $t$-$\mathrm{Zr_2O_3}$ yielded a (110) peak at $40.2^{\mathrm{o}}$ and a (112) peak at $51.6^{\mathrm{o}}$.

In conclusion, we have predicted tetragonal semi-metallic $\mathrm{Hf_2O_3}$ and $\mathrm{Zr_2O_3}$ structures as the ground state highly oxygen deficient hafnia and zirconia which undergo a monoclinic-to-tetragonal phase transition. Their semi-metallic properties are characterized by an overlap of the valence band maximum and conduction band minimum at different points of the Brillouin zone, and by the low density of conduction electrons and holes. Also, $t$-$\mathrm{Hf_2O_3}$ may be the physical origin of the conductive state in hafnium-based RRAM.

This work is financially supported by the Nanosciences Foundation of Grenoble (France) in the frame of the Chairs of Excellence awarded to L.R.C. Fonseca in 2008 and to Y. Nishi in 2010. LRCF also acknowledges CNPq for financial support. The calculations were performed on the Stanford NNIN (National Nanotechnology Infrastructure Network) Computing Facility funded by the National Science Foundation of USA. We specially thank Dr. Blanka Magyari-K\"ope from Stanford University for pointing out the experimental results in Ref.~\onlinecite{Osaka}.

\end{document}